\documentclass[sigconf]{acmart}
\usepackage{booktabs}
\usepackage{subfig}
\usepackage{amsmath,balance}

\AtBeginDocument{%
  }

\copyrightyear{2022}
\acmYear{2022}
\setcopyright{rightsretained}
\acmConference[ISLPED '22]{ACM/IEEE International Symposium on Low Power Electronics and Design}{August 1--3, 2022}{Boston, MA, USA}
\acmBooktitle{ACM/IEEE International Symposium on Low Power Electronics and Design (ISLPED '22), August 1--3, 2022, Boston, MA, USA}
\acmDOI{10.1145/3531437.3539699}
\acmISBN{978-1-4503-9354-6/22/08}

\newcommand{\esnote}[1]{{\em{\color{red}  [Elaheh's note: #1]}}}
\newcommand{\jynote}[1]{{\em{\color{blue}  [Jingyao's note: #1]}}}

\begin{document}

\title{Sealer: In-SRAM AES for High-Performance and Low-Overhead Memory Encryption}

\author{Jingyao Zhang}
\affiliation{%
  \institution{Department of Computer Science\\ University of California, Riverside}
  \country{USA}
  }
\email{jzhan502@ucr.edu}

\author{Hoda Naghibijouybari}
\affiliation{%
  \institution{Department of Computer Science\\ Binghamton University}
  \country{USA}
  }
\email{hnaghibi@binghamton.edu}

\author{Elaheh Sadredini}
\affiliation{%
  \institution{Department of Computer Science\\ University of California, Riverside}
  \country{USA}
  }
\email{elaheh@cs.ucr.edu}

\renewcommand{\shortauthors}{Zhang et al.}

\begin{abstract}
To provide data and code confidentiality and reduce the risk of information leak from memory or memory bus, computing systems are enhanced with encryption and decryption engine.   
Despite massive efforts in designing hardware enhancements for data and code protection, existing solutions incur significant performance overhead as the encryption/decryption is on the critical path.
In this paper, we present \textit{Sealer}, a high-performance and low-overhead in-SRAM memory encryption engine by exploiting the massive parallelism and bitline computational capability of SRAM subarrays. \textit{Sealer} encrypts data before sending it off-chip and decrypts it upon receiving the memory blocks, thus, providing data confidentiality. 
Our proposed solution requires only minimal modifications to the existing SRAM peripheral circuitry. 
\textit{Sealer} can achieve up to two orders of magnitude throughput-per-area improvement while consuming 3$\times$ less energy compared to prior solutions. 
\end{abstract}



\maketitle

\section{Introduction}
\vspace{-0.1cm}





Healthcare organizations, businesses, and governments rely heavily on secure computer systems for their daily activities and business conduct.
To provide data confidentiality, a wide range of computational devices, from high-end servers to low-power IoT devices, implement data encryption standards. 
In these devices, only the processor chip is considered secure and trusted hardware in the system \cite{costanIntelSgxExplained2016}. Anything outside the processor chip boundary is typically assumed vulnerable and untrusted. In such a threat model, any data sent off-chip (i.e., to the memory or on the cloud) is potentially at risk of being manipulated, tampered with, or leaked~\cite{haldermanLestWeRemember2009b}. Novel memory technologies, such as Non-Volatile Memories (NVMs), can retain data for a long time after power loss, thus, making memory protection even more critical. 

To protect against memory and bus attacks (e.g., bus snooping and cold boot attacks \cite{haldermanLestWeRemember2009b}), processor vendors are increasingly adding support for protecting the integrity and confidentiality of data through integrity verification and encryption on the on-chip memory controller \cite{gueronMemoryEncryptionEngine2016b,kaplanAMDMemoryEncryption2016a}.
Memory encryption typically uses Advanced Encryption Standard (AES) block cipher to encrypt plaintext going to or decrypt ciphertext from the main memory, using a key re-generated on each system reboot. However, real-time encryption or decryption at every memory access in the critical path results in severe performance overhead \cite{mathew53GbpsNativeGF2010}. 

To address the performance issue, recent prior works have proposed in-memory and near-memory encryption/decryption solutions~\cite{xieSecuringEmergingNonvolatile2018,agaInvisiMemSmartMemory2017,angiziPIMALogicNovelProcessinginmemory2018}. 
Xie et al.~\cite{xieSecuringEmergingNonvolatile2018} propose AIM, an in-memory AES engine that provides bulk encryption of data blocks in NVM for mobile devices, and encryption is executed only when the device is shut down or put into sleep/screen-lock mode. Despite its computational efficiency, AIM does not protect data confidentiality against bus and memory attacks. Aga et al.~\cite{agaInvisiMemSmartMemory2017} present InvisiMem, which uses the logic layer in 3D stacked memory to implement cryptographic primitives.
However, the InvisiMem design expands the trusted computing base (TCB) to the logic layer of the memory. To protect data confidentiality against physical attacks on both memory and bus, an efficient real-time encryption/decryption engine needs to be implemented on the processor chip.

To provide a low-cost, and real-time on-chip encryption engine, for the first time, this paper proposes to re-purpose 6T SRAM subarrays into active large vector computational units to perform encryption and decryption on-chip. Our solution, \textit{Sealer}, exploits intrinsic parallelism and bitline computational capability of memory subarrays for fast and low-overhead AES implementation, and incurs only a negligible area overhead (less than 1.55\%) compared to the traditional SRAM arrays. 
\textit{Sealer} provides the same level of memory confidentiality as Intel Memory Encryption Engine (MEE)~\cite{gueronMemoryEncryptionEngine2016b} or AMD Secure Memory Encryption (SME)~\cite{kaplanAMDMemoryEncryption2016a}, and unlike InvisiMem, does not extend the TCB.
\textit{Sealer} can act as conventional SRAM when the encryption/decryption is unnecessary.

In summary, the paper makes the following \textbf{contributions:}
\begin{itemize}
    \item We present \textit{Sealer}, a real-time in-SRAM AES engine to provide data confidentiality by encrypting the plaintext on the CPU chip.
    Our proposed architecture effectively stores the required data 
    for encryption into the same 
    subarray. This allows the peripherals and resources to be shared among different computation and communication units, thus, reducing performance and hardware overhead. 
    \item We present an algorithm and architecture methodology to efficiently map the AES algorithm to the \emph{Sealer} architecture. By fusing the \textit{SubBytes} and \textit{ShiftRows} stages in AES, we can interleave the computation at a finer granularity and maximally exploit on-the-fly computation to significantly reduce data movement and \textit{write} cycles.
    \item We compare \textit{Sealer} with several on-chip and in-memory AES encryption engines and show that our solution has up to 323$\times$ performance improvement, up to 91$\times$ throughput-per-area improvement, and 3$\times$ lower energy consumption compared to prior solutions. To separate the architectural and technology contribution of \textit{Sealer}, we evaluate an in-NVM AES engine on SRAM (\emph{AIM-SRAM}) and find that our solution achieves 6$\times$ higher performance than AIM-SRAM due to architectural contribution and 18$\times$ better performance due to technology benefits.
\end{itemize}

\vspace{-0.3cm}
\section{Background and Threat Model}
\vspace{-0.1cm}
\subsection{Advanced Encryption Standard}
\begin{figure}[htp]
\vspace{-0.3cm}
\centerline{\includegraphics[width=2.5in]{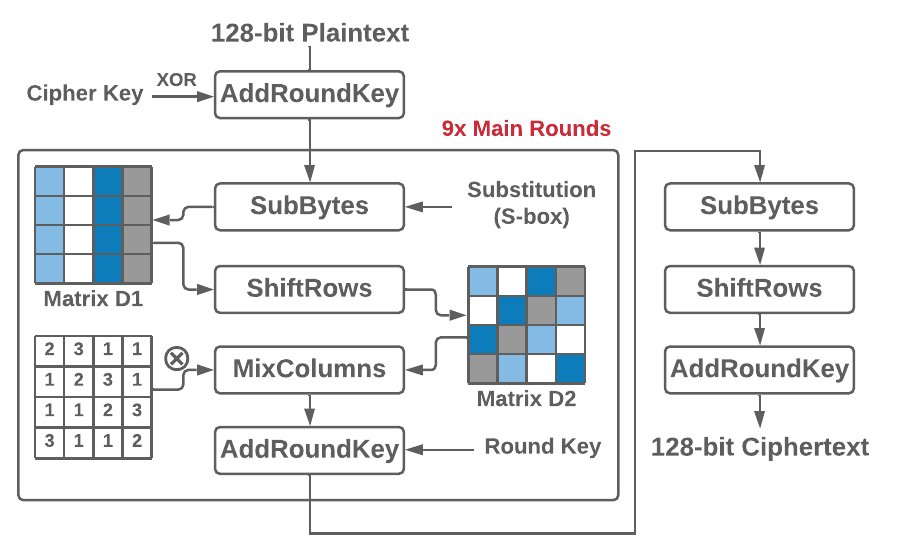}}
\vspace{-0.3cm}
\caption{The steps of AES for a 128-bit plaintext.
}
\label{AES_flow}
\vspace{-0.5cm}
\end{figure}

\vspace{-0.1cm}
The Advanced Encryption Standard (AES) in cryptography, also known as Rijndael cryptography, is a block encryption standard.
The AES encryption process operates on a 4×4 matrix of bytes
whose initial value is a plaintext block (the element size in the matrix is one byte). 
Each AES encryption round (except the last one) consists of four steps where the output of each stage is used as an input for the next stage and described as follows.  
(1) \emph{AddRoundKey}: each byte in the matrix is XORed with the round key. Each round key is generated by the key generation scheme from a given cipher key. 
(2) \emph{SubBytes}: each byte is substituted with its corresponding byte in the S-box block using a nonlinear substitution function commonly implemented with lookup tables (LUTs). 
(3) \emph{ShiftRows}: a round-robin shift is performed for each row in the matrix. The first row is unchanged while each element of the second row are shifted to the left by 1 byte. Then for the third and fourth row, elements are shifted to the left by 2 and 3 bytes, respectively (see matrix D1 to D2 transformation in Figure \ref{AES_flow}). 
(4) \emph{Mixcolumns}: this step uses a linear transformation to fully mix the four bytes of each column. In AES, the \emph{MixColumns} stage is omitted from the last encryption loop and replaced with another \emph{AddRoundKey}.

The overall process of AES for a 128-bit plaintext is shown in Figure \ref{AES_flow}. First, the plaintext is XORed with the original cipher key for initialization. Then after 9 main rounds, each of which consists of \emph{AddRoundKey}, \emph{SubBytes}, \emph{ShiftRows}, and \emph{MixColumns}, and the final round without \emph{MixColumns}, the 128-bit ciphertext is generated.

\vspace{-0.3cm}
\subsection{Computing in SRAM}
\begin{figure}[htp]
\vspace{-0.3cm}
\centerline{\includegraphics[width=1.7in]{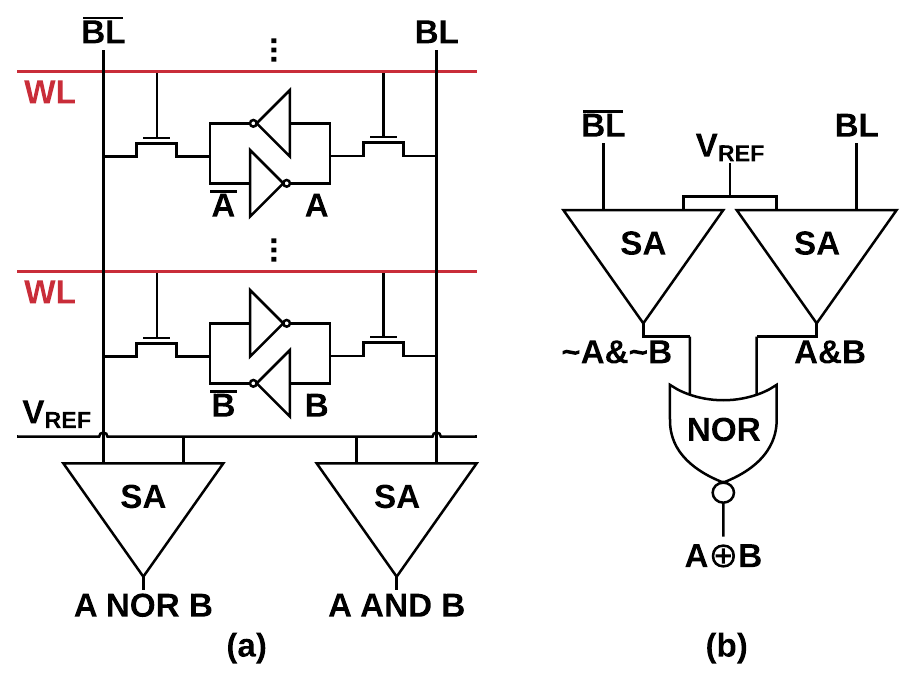}}
\vspace{-0.3cm}
\caption{The XOR bitline operation using 6T SRAM cells.}
\label{bitline}
\vspace{-0.5cm}
\end{figure}

\vspace{-0.1cm}
In-SRAM computing relies on the underlying bitline computation by activating more than one row in the SRAM subarray \cite{jeloka28NmConfigurable2016a,sadredini2020impala}. 
The bitwise AND and NOR operations in SRAM are implemented directly by utilizing sense amplifiers (SAs) with multiple wordlines activated, as shown in Figure \ref{bitline}(a). 
The SA on the bitline ($BL$) can sense a voltage higher than $V_{ref}$ only if all the cells in the activated rows connected to the corresponding $BL$ contain `1'. 
This means the SA will sense `1', thus, achieving element-wise AND operation.
The SA on the bitline-bar ($\overline{BL}$) will sense a voltage higher than $V_{ref}$ only if all the cells in the activated rows connected to the corresponding $\overline{BL}$ contain `1', which, in turn, implies that all the cells in the activated rows connected to the corresponding $BL$ contain '0'.  
This means the SA will sense `1', thus, achieving element-wise NOR operation.
Using the logical bitwise AND and NOR operations, the XOR operation can be performed, as shown in Figure \ref{bitline}(b).



Many studies based on computing in SRAM have been proposed \cite{agaComputeCaches2017,eckertNeuralCacheBitserial2018,fujikiDualityCacheData2019,subramaniyanCacheAutomaton2017}.
Cache Automaton uses a sense-amplifier cycling technique to read out multiple bits in one time slot, thus significantly reducing input symbol match time \cite{subramaniyanCacheAutomaton2017}.
Based on the described NOR, AND, and XOR operations, Compute Cache extends the logical operations by slightly modifying the SA design in \cite{agaComputeCaches2017}.
In this paper, we utilize the XOR functionality presented in \cite{agaComputeCaches2017}, and slightly modify it (similar to \cite{gencache}) to be able to perform the shift operation (required for the proposed fused \emph{ShiftRows} and \emph{MixColumns} stages) in place without the need to store the intermediate result back into the subarray, thus, reducing processing cycles.  

\vspace{-0.2cm}
\subsection{Threat Model}\label{sec:threatmodel}
\vspace{-0.1cm}
Similar to the state-of-the-art secure memory architectures~\cite{gueronMemoryEncryptionEngine2016b}, we assume the only trusted component is the processor chip and we do not expand the TCB of the system. Therefore, bus and memory are vulnerable and untrusted. We assume the attacker has physical access to the system and can snoop on the bus, or scan the memory.
This work proposes high-performance encryption for protecting data confidentiality. Addressing timing channels, replay attacks, and access pattern leakage threat models are left for future work.

\vspace{-0.2cm}
\section{Related Work}
\vspace{-0.1cm}




Hämäläinen et al.\cite{hamalainenDesignImplementationLowarea2006} (EE-1) and Mathew et al.\cite{mathew53GbpsNativeGF2010} (EE-2) propose on-die AES encryption engines.
EE-2 computes the entire AES round in composite-field arithmetic, resulting in delay reduction in the critical S-box unit. They also use folded datapath design to reduce the wiring complexity of \emph{ShiftRows} permutations. 
Wang et al. propose DW-AES\cite{wangDWAESDomainwallNanowirebased2016} by utilizing the domain wall nanowire to implement the AES encryption algorithm in NVM. 
Though the throughput of DW-AES is 3.6$\times$ higher than the CMOS ASIC proposed in \cite{mathew340MV2892015}, the latency for one data block is still not low enough to accommodate real-time encryption. 

To improve the performance of memory encryption, recent works propose in-memory encryption architectures. Xie et al.\cite{xieSecuringEmergingNonvolatile2018} present AIM, an efficient in-NVM implementation of AES. AIM provides bulk encryption/decryption for mobile devices to protect memory content only when the device is shut down, locked, or in sleep mode. 
Therefore, AIM does not provide real-time memory encryption to protect against bus and memory attacks.
Aga et al. propose InvisiMem \cite{agaInvisiMemSmartMemory2017}, which expands the trust base to the logic layer of 3D stacked memory to implement encryption. InvisiMem guarantees confidentiality, integrity, and protecting access patterns using a packetized interface and authentication to establish a secure communication channel between processor and memory. \emph{Sealer} does not expand the TCB of the system and is the first work that proposes in-SRAM encryption for data confidentiality. 

Commercial examples of memory protection solutions are Memory Encryption Engine (MEE)~\cite{gueronMemoryEncryptionEngine2016b} in Intel SGX \cite{costanIntelSgxExplained2016} and AMD's Secure Memory Engine (SME)~\cite{kaplanAMDMemoryEncryption2016a}. 
The hardware component of SGX (i.e., MEE) protects the confidentiality, integrity, and freshness of processor-DRAM traffic for secure enclaves. The confidentiality protection in MME and SME are implemented by AES-128 encryption. 
\emph{Sealer} provides the MEE- and SME-equivalent guarantee for memory confidentiality, but with higher performance and lower area overhead by exploiting intrinsic parallelism and bitline computational capability of SRAM subarrays.

We compare the performance of \emph{Sealer} with EE-1  ~\cite{hamalainenDesignImplementationLowarea2006}, EE-2~\cite{mathew53GbpsNativeGF2010}, DW-AES~\cite{wangDWAESDomainwallNanowirebased2016}, and AIM \cite{xieSecuringEmergingNonvolatile2018} in Section~\ref{sec:evaluation}.  
We are not able to compare the performance of \emph{Sealer} with Intel's MEE, AMD's SME, and Invisimem mainly because these designs provide other memory protections, such as data integrity. Therefore, we cannot isolate the encryption performance from other components in their designs. 
\vspace{-0.4cm}
\section{Implementation}
\vspace{-0.1cm}
\subsection{Data Organization}
\vspace{-0.1cm}

\begin{figure}[htp]
\vspace{-0.3cm}
\centerline{\includegraphics[width=3.2in]{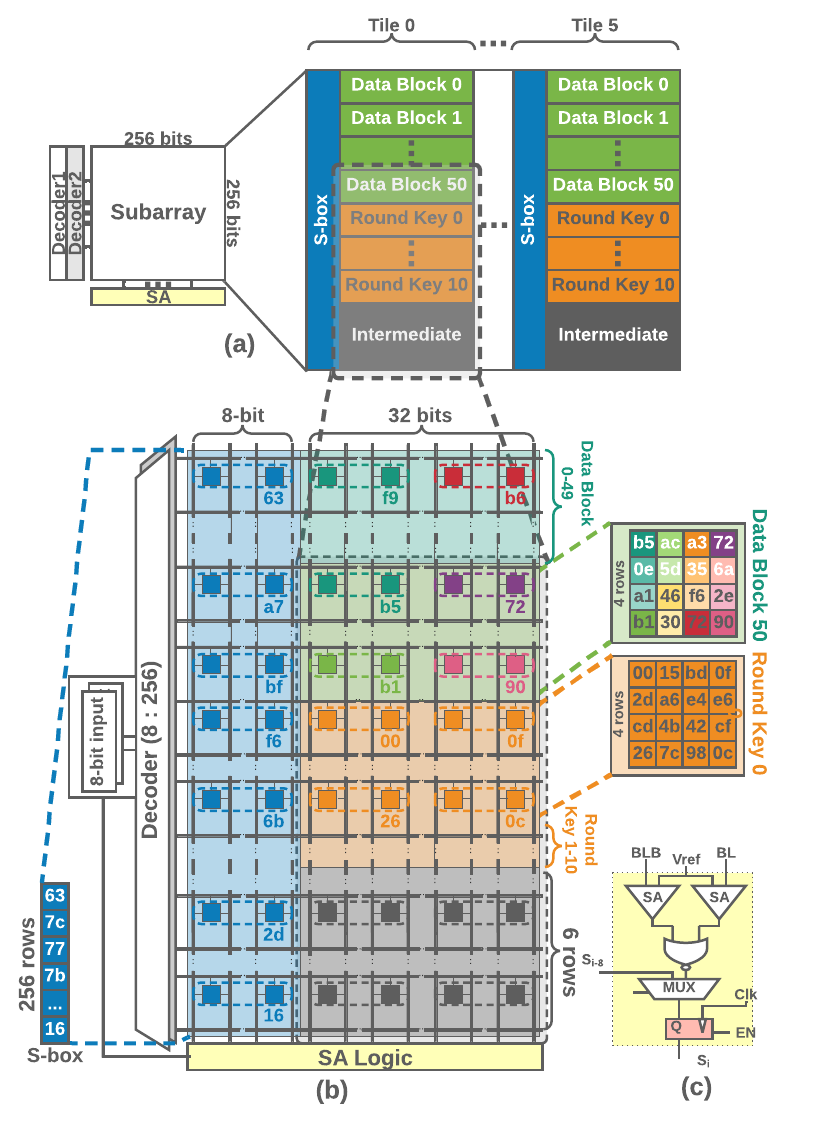}}
\vspace{-0.5cm}
\caption{(a) Data organization in one subarray. (b) The detailed structure of the first S-box and data (Tile 0) with three inputs of AES encryption algorithm: S-box (256-row, 8-bit), 128-bit plaintext and 128-bit cipher key. (c) The structure of the sense amplifier to support XOR and shift operations.}
\label{overall}
\vspace{-0.9cm}
\end{figure}

In this paper, an efficient implementation of the AES algorithm is accomplished using bitline computing of SRAM arrays, and can be realized by re-purposing a portion of the L3 cache or by replacing the existing on-chip encryption hardware. 
To efficiently utilize bitline computing, we organize the S-box, plaintext, keys, and intermediate data required by the AES algorithm in the same subarray, as shown in Figure \ref{overall}(a).
We assume $256\times256$ SRAM subarrays, following Intel’s SandyBridge L3 cache structure \cite{lempel2ndGenerationIntel2011}. 
As depicted in Figure \ref{overall}, each subarray consists of 6 \emph{Tiles}, where each \emph{Tile} can store 51 data blocks ($Row[0]-Row[203]$ - each data block requires four rows), one set of round keys ($Row[204]-Row[247]$), and intermediate data required for \textit{MixColumns} stage ($Row[248]-Row[253]$).
The first 8 columns of the \emph{Tile} are used to hold the S-box.  
All the data blocks arranged in the same set of rows can encrypt/decrypt data in parallel (i.e., 6 data blocks per subarray). Data blocks in the same \emph{Tile} cannot be computed parallelly due to the shared bitlines.
 
The S-box is originally a 16$\times$16 matrix where each element is an 8-bit data. 
To avoid adding the extra LUT logic in the SRAM structure for S-box substitution in the \emph{SubBytes} stage, and also to eliminate the communication and wiring overhead between the data block subarrays and the LUTs, \emph{Sealer} proposes to re-organize the S-box into a matrix of 256 rows where each row has an 8-bit element (as shown in Figure \ref{overall}(b) - the blue matrix) and utilize the same subarray as the plaintext and round keys are located. The memory decoder is used to decode the 8-bit input data to select the corresponding element for the substitution. This brings two advantages in \emph{Sealer} compared to the prior work, AIM \cite{xieSecuringEmergingNonvolatile2018}, from the architectural aspect; (1) all the required data for performing the encryption is within the same subarray, and this brings the opportunity to fuse computation in different stages while the data is read into SA, and also perform intermediate data generation on-the-fly without the need to write them back into the subarray, thus, reducing the total number of cycles, and (2) unlike AIM that requires additional modules, such as LUTs and multiplexers, we repurpose the existing resources (i.e., memory array and peripherals) to perform the required computation, and thus, avoiding extra hardware overhead. Details are discussed in Sections \ref{subbytes} and \ref{mixColumns}.


\begin{figure*}[htp]
\vspace{-0.0cm}
\centerline{\includegraphics[width=6.5in]{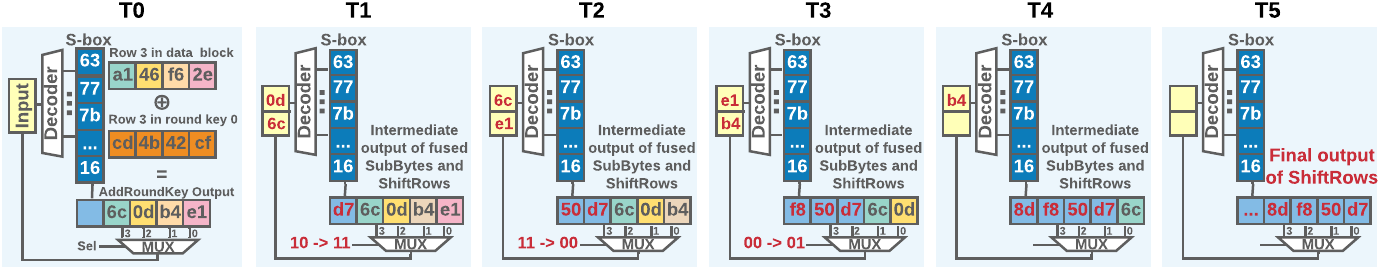}}
\vspace{-0.3cm}
\caption{Fusion of the \emph{SubBytes} and \emph{ShiftRows} stages for the third row of the data matrix.}
\label{fused}
\vspace{-0.5cm}
\end{figure*} 

\vspace{-0.3cm}
\subsection{AddRoundKey}
\vspace{-0.1cm}

\emph{AddRoundKey} receives two matrices, the data matrix (i.e., the plaintext) and the key matrix (i.e., the round key).
Each corresponding byte in the data matrix is bit-wise $\text{XOR}$ed with each byte in the key matrix by activating two rows using two decoders.
We follow the bitline XOR functionality implementation in \cite{xieSecuringEmergingNonvolatile2018}, which costs 3 times longer than a single subarray read/write access.

As depicted in Figure \ref{overall} (b), the data matrix (green area) and the key matrices (orange area) rows are arranged in consecutive rows (column-aligned) in the subarray, and the elements within the same row in the matrix are located next to each other horizontally in the subarray (i.e., each 128-bit data matrix or 128-bit round key requires 4 consecutive rows and 32 consecutive columns). 
This means that \emph{Sealer} can perform the bitwise $\text{XOR}$ in parallel on 192 bits (4 elements in one row of a \emph{Tile}, across 6 \emph{Tiles} in a subarray) of plaintext within one subarray in only 3 cycles (note that several subarrays can work all in parallel without almost any performance overhead!).
The \emph{AddRoundKey} output (i.e., the bitwise XOR) is sensed in the SAs (Figure \ref{overall} (c)) by activating the equivalent rows in the data matrix and key matrix. 
To maximally utilize the data retained in the sense amplifier (i.e., row buffer hit) and eliminate unnecessary write cycles, we perform the computation of the next stage on the output of the first row before writing it back to subarray.





\vspace{-10pt}

\subsection{Fused SubBytes and ShiftRows} \label{subbytes}
\vspace{-0.1cm}
To efficiently\enlargethispage{5pt} implement the \emph{SubBytes} and \emph{ShiftRows} stages and reduce the total number of processing cycles, we fuse the \emph{ShiftRows} and \emph{SubBytes} stages by selecting the right order of the elements from the output of \emph{AddRoundKey} and feeding the element as the input address of the decoder to read the substitution value. The order is determined by the number of shifts required in the \emph{ShiftRows} stage.

AES defines a substitution box (S-box), which consists of a $16\times16$ byte matrix and is used to substitute each byte of the data block with the corresponding byte in the S-box matrix by using the four MSB bits to select one of the 16 rows and the four LSB bits to choose one of the 16 columns. 
Instead of adding LUTs and incurring the extra hardware overhead, \emph{Sealer} re-purposes the existing resources in SRAM and stores the S-box in the same subarray as data block and keys are stored. We accordingly reorganize the conventional $16\times16$ byte matrix to a matrix of $256\times1$ where each row represents one byte (Figure \ref{overall} - blue S-box matrix). 
We then use the result of \emph{AddRoundKey} stage, which is already stored in SAs, as the input of the decoder and read the substitution bytes consecutively.
The input of the decoder is equipped with a two 8-bit entry FIFO buffer, to enable the in-place substitution (\emph{SubBytes}) and shifting (\emph{ShiftRows}). 

Figure \ref{fused} demonstrates how the proposed \emph{fused SubBytes and ShiftRows} stage works for the third row in the data bock matrix using a step-by-step example. 
In the initial time step (T0), the XOR results of the third row in the data matrix (a1, 46, f6, 2e) and the third row of the round key matrix (cd, 4b, 42, cf) are written into the SA buffers (6c, 0d, b4, e1). 
We first explain the conventional computation in prior work (i.e., consecutive \emph{SubBytes} and \emph{ShiftRows}), and then describe the proposed fused approach in \emph{Sealer}. 

Assume that the substitution bytes (i.e., the output of the \emph{SubBytes} stage) for (6C, 0d, b4, e1) are (50, d7, 8d, f8). The output of the \emph{SubBytes} will be the input of \emph{ShiftRows} in the next stage. Because the computation is performed on the third row, the \emph{ShiftRows} stage will shift the byte to the left two times using cyclic shift, i.e., the output of the \emph{ShiftRows} stage will be (8d, f8, 50, d7).

In the proposed fused approach, we first select the element that its corresponding substitution byte will be located on the right-most side of the array after the \emph{ShiftRows} stage is completed, i.e., $sel = 10 (``0d")$ will be read into the input buffer. Then, we select the elements in the \emph{AddRoundKey} output one by one according to their corresponding locations in the \emph{ShiftRows} output from the right-most side element to the left-most side element in the array, i.e., $sel = 10 (0d) \to 11 (6c)\to 00 (e1) \to 01 (b4)$.

For the correct functionality, the input buffer should have two entries. In T1, 0d and 6c are read to the FIFO buffer. Then, the decoder takes 0d as the input address and the corresponding substitution byte is read into the row buffer (i.e., d7). In T2, first, e1 is selected (Sel = 00) and read into the second entry of the input buffer. Then, the intermediate output is shifted to the right by one byte, and finally, the decoder takes 6c as the input and the corresponding substitution byte is read into the row buffer (i.e., 50).
In T3, first, b4 is selected (Sel = 01) and read into the second entry of the input buffer. Then, the intermediate output is shifted to the right by one byte, and finally, the decoder takes e1 as the input and the corresponding substitution byte is read into the row buffer (i.e., f8).
In T4, first, the intermediate output is shifted to the right by one byte, and then, the decoder takes b4 as the input and the corresponding substitution byte is read into the row buffer (i.e., 8d). 
In T5, the intermediate buffer is shifted to the right by one byte and the final output 
is generated and ready to be used by the \emph{MixColumns} stage.

The shift operation is implemented by introducing a latch and a multiplexer within the SA \cite{gencache}, as shown in Figure \ref{overall}(c).
The proposed fusion approach greatly reduces write cycles, thus reducing processing latency.
It is important to note that the fused \emph{SubBytes} and \emph{ShiftRows} computation are done in parallel across all the \emph{Tiles} (i.e., one data block in each \emph{Tile} as the data blocks within one \emph{Tile} cannot be processed in parallel) in and across subarrays.  

\vspace{-10pt}

\subsection{MixColumns} \label{mixColumns}
\vspace{-0.1cm}

In\enlargethispage{5pt} the \emph{MixColumns} stage, each column of the input data matrix (i.e., the output matrix of \emph{ShiftRows} stage) is multiplied with a two-dimensional constant array, called the fixed matrix, to obtain the corresponding output column.
All the addition and multiplication are both defined over a finite field. 
To achieve efficient computation in SRAM, and to reduce the resource consumption and the number of SRAM accesses for intermediate results, we decompose the matrix multiplication of data columns and the fixed matrix into matrix elements, an intermediate value $T_c$, and the product of matrix elements, following \cite{xieSecuringEmergingNonvolatile2018}.
The decomposition equation is as follows: 
\begin{equation}\label{equ:matrix}
    \scalebox{0.72}{%
\ensuremath{
\left[\begin{matrix}
2 & 3 & 1 & 1\\
1 & 2 & 3 & 1\\
1 & 1 & 2 & 3\\
3 & 1 & 1 & 2
\end{matrix}\right]
\left[\begin{matrix}
B_{0,c} \\
B_{1,c} \\
B_{2,c} \\
B_{3,c} 
\end{matrix}\right]
 =
\left[\begin{matrix}
2 B_{0,c}+3 B_{1,c}+ B_{2,c}+ B_{3,c} \\
B_{0,c}+2 B_{1,c}+3 B_{2,c}+ B_{3,c} \\
B_{0,c}+B_{1,c}+2 B_{2,c}+3 B_{3,c} \\
3 B_{0,c}+B_{1,c}+ B_{2,c}+2 B_{3,c} 
\end{matrix}\right]
\\
= 
\left[\begin{matrix}
T_c \oplus 2 B_{0,c} \oplus 2 B_{1,c} \oplus B_{0,c} \\
T_c \oplus 2 B_{1,c} \oplus 2 B_{2,c} \oplus B_{1,c} \\
T_c \oplus 2 B_{2,c} \oplus 2 B_{3,c} \oplus B_{2,c} \\
T_c \oplus 2 B_{0,c} \oplus 2 B_{3,c} \oplus B_{3,c}, 
\end{matrix}\right]}
        }
\end{equation}
where $B_{r,c}$ denotes the byte in row $r$ and column $c$, and $T_c$ is the intermediate result for column $c$. $T_c$ is calculated as: 
\begin{equation}
    T_c=B_{0,c}\oplus B_{1,c} \oplus B_{2,c} \oplus B_{3,c}.
\end{equation}

\begin{figure}[tp]
\vspace{-0.3cm}
\centerline{\includegraphics[width=3.3in]{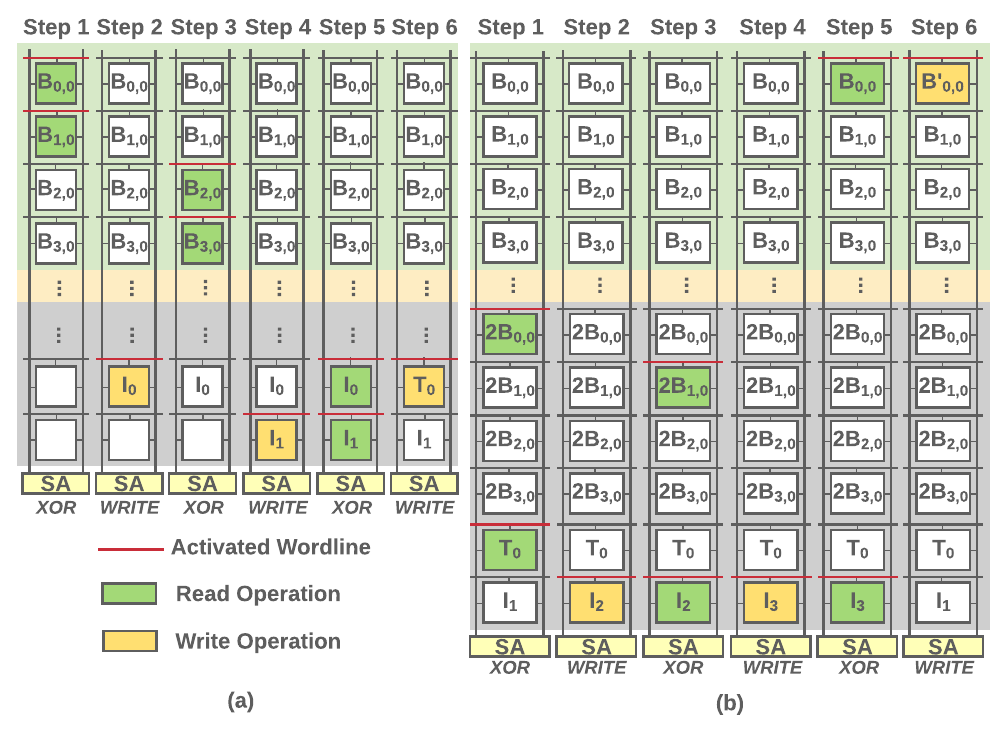}}
\vspace{-0.3cm}
\caption{Example of \emph{MixColumns} stage.
}
\label{mixcolumn}
\vspace{-0.6cm}
\end{figure}

The computation of $2\times B_{r,c}$ elements in Equation \ref{equ:matrix} is done in \emph{ShiftRows} stage by shifting the output of \emph{ShiftRows} stage to the left by one bit and storing them back in the intermediate region of the subarray. 
Figure \ref{mixcolumn}(a) shows how $T_c$ in each column is calculated and stored in the subarray in 6 steps using an example (note that the calculation of $T_c$ is fully parallel in each column). 
To calculate $T_0$, in Step 1, the two (red) worldlines are activated simultaneously to calculate the XOR of $B_{0,0}$ and $B_{1,0}$, and store the result ($I_0$) in the intermediate region in Step 2. Next, the XOR of $B_{2,0}$ and $B_{3,0}$ is calculated in Step 3 and written back to the array ($I_1$) in Step 4. Finally, $T_0$ is calculated (Step 5) and written back (Step 6). 
Figure \ref{mixcolumn}(b) shows how the final output is calculated for $T_c \oplus 2 B_{0,c} \oplus 2 B_{1,c} \oplus B_{0,c}$ by activating the corresponding rows, calculating the XOR of the activated cells, and writing the intermediate results back. After 6 steps, the output of the \emph{MixColumn} ($B^{'}_{0,0}$) overwrites input ($B_{0,0}$).

\vspace{-0.15cm}
\section{Key generation and Storage}
\vspace{-0.1cm}
Similar to state-of-the-art memory encryption engines~\cite{gueronMemoryEncryptionEngine2016b}, we assume the encryption key is generated using a hardware random number generator and implemented in the processor chip. The key is inaccessible outside of the chip.
The expansion of the given cipher key can also be utilized to obtain 11 partial keys, which are used in the initial round, the 9 main rounds, and the final round.
The expansion process can be implemented in the subarray, including shift operations between columns, the \emph{SubBytes} stage, the XOR operations between rows and a given reconstruction matrix.
The generated keys will be stored within the subarray (Figure \ref{overall}).



\vspace{-0.2cm}
\section{Evaluation}\label{sec:evaluation}
\vspace{-0.1cm}

\subsection{Evaluation Methodology}



In this section, we evaluate the performance, throughput-per-area and energy/power of \emph{Sealer} 
and compare it to prior in-memory encryption solutions \cite{xieSecuringEmergingNonvolatile2018,wangDWAESDomainwallNanowirebased2016} and on-chip memory encryption implementations \cite{hamalainenDesignImplementationLowarea2006,mathew53GbpsNativeGF2010}.
AIM architecture\cite{xieSecuringEmergingNonvolatile2018} is implemented in-MRAM (\emph{AIM-NVM}) and \emph{Sealer} utilizes
SRAM-based subarrays. To decouple architectural contribution from the technology contribution and have an apples-to-apples comparison between AIM and \emph{Sealer}, we model and evaluate the architecture of AIM in SRAM (\textit{AIM-SRAM}).
The area overhead evaluation presented in the AIM is based on relative numbers. To directly compare the area overhead of AIM-NVM and AIM-SRAM with \emph{Sealer}, we use NVSim \cite{dongNVSimCircuitlevelPerformance2012} to obtain the absolute area parameters for the peripherals used in AIM-NVM. 
The read/write access latency to a $256\times256$ 6T SRAM subarray is 163 $ps$, and 
an XOR operation in SRAM costs 489 $ps$, which are extracted from the SPICE simulations with 28nm SOI CMOS process \cite{subramaniyanCacheAutomaton2017,agaComputeCaches2017}. 
DESTINY \cite{destiny} is used to evaluate the energy and power consumption of SRAM-based and NVM-based designs.

The baselines are (1) EE-1 \cite{hamalainenDesignImplementationLowarea2006}, which introduces an AES encryption based on a dedicated engine for low power consumption at 290 MHz, (2) EE-2 \cite{mathew53GbpsNativeGF2010}, which presents a high-frequency integrated circuit encryption engine at up to 2.13 GHz, (3) DW-AES \cite{wangDWAESDomainwallNanowirebased2016}, which uses domain-wall nanowires to implement AES encryption in NVM at 30MHz, (4) AIM-NVM, and (5) AIM-SRAM.



\vspace{-0.2cm}
\subsection{Latency Analysis} \label{PerformanceAnalysis}
\begin{figure}[htp]
\vspace{-0.3cm}
\centerline{
\includegraphics[width=3.1in]{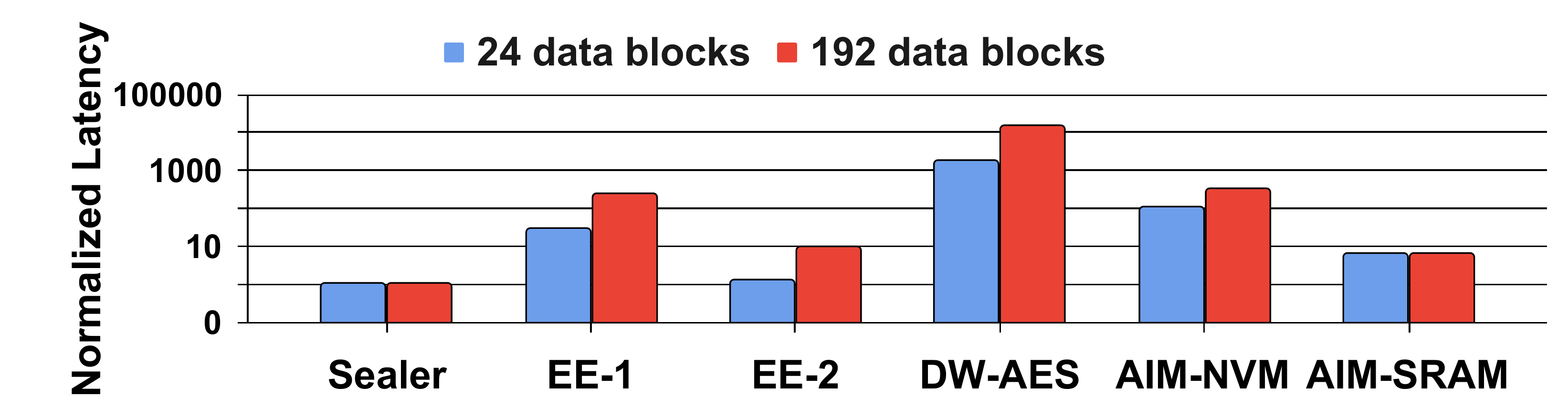}
}
\vspace{-0.3cm}
\caption{Latency comparison among different baselines normalized to \emph{Sealer} for encrypting 24 and 192 data blocks.}
\label{combine}
\vspace{-0.4cm}
\end{figure}

\vspace{-0.1cm}
Figure \ref{combine} compares the data encryption latency of different solutions normalized to the \emph{Sealer} latency for 24 data blocks (384 bytes) and for 192 data blocks (3072 bytes, which is approximately the size of one cache slice in L3 \cite{lempel2ndGenerationIntel2011}). 
\emph{Sealer} has 30$\times$ (243$\times$), 1.22$\times$ (9.8$\times$), and 1880$\times$ (15040$\times$) lower latency than EE-1, EE-2, and DW-AES, respectively, for encrypting 24 data blocks (192 data blocks).
For all these solutions, \emph{Sealer's} performance improves significantly when increasing the memory capacity because of the intrinsic bit-level and subarray-level parallelism in memory. 
For example, EE-2 can operate on only four blocks in parallel, 
while \emph{Sealer} can encrypt 192 data blocks simultaneously when utilizing the 2MB SRAM. 
In general, in-memory solutions (\emph{Sealer}, AIM-NVM, and AIM-SRAM) provide higher parallelism than dedicated hardware engines.  

\begin{figure}[htp]
\vspace{-0.3cm}
\centerline{
\includegraphics[width=2.2in]{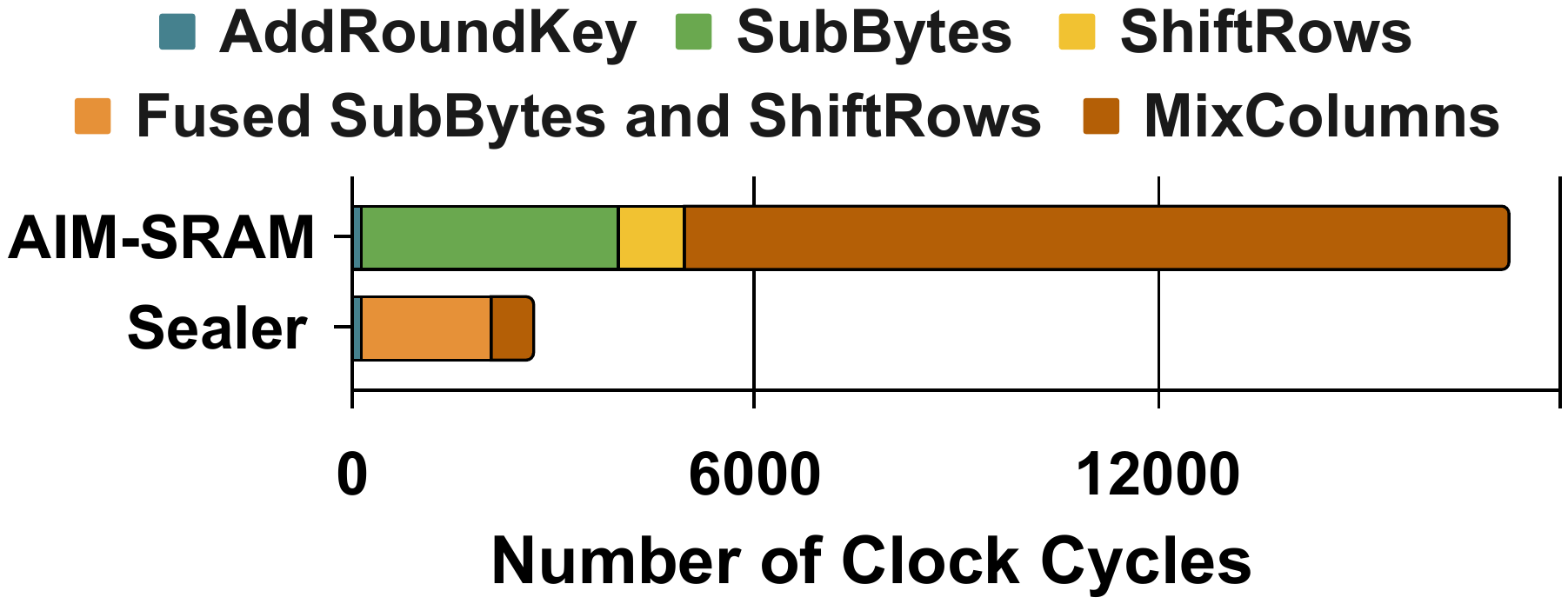}
}
\vspace{-0.3cm}
\caption{Breakdown of processing cycles in AIM-SRAM and \emph{Sealer} for encrypting 6 data blocks.}
\label{cyclenumber}
\vspace{-0.3cm}
\end{figure}

\emph{Sealer} has almost 6.5$\times$ lower latency than AIM-SRAM (architectural contribution of \emph{Sealer} compared to AIM). 
This is because the required number of cycles for encrypting six data blocks in a subarray in AIM-SRAM is almost 6.5$\times$ higher than \emph{Sealer}, as shown in Figure \ref{cyclenumber}. 
Figure \ref{cyclenumber} compares the cycle breakdown for different stages in \emph{Sealer} and AIM. \emph{AddRoundKey} stage has the same number of cycles in both architectures. However, the fused \emph{SubBytes} and \emph{ShiftRows} stage in \emph{Sealer} requires 59.5\% fewer cycles than the total cycles for \emph{SubBytes} and \emph{ShiftRows} stages in AIM. This benefit is enabled by storing all required data (i.e., data blocks, S-box, round keys, and intermediate data) in the same subarray, which facilitates computation fusion among different stages, thus, resulting in fewer write cycles. 
Moreover, in the \emph{MixColumns} stage, unlike AIM which uses additional LUTs, \emph{Sealer} utilizes the existing rows and peripherals in the under-processed subarray to store intermediate results, thus, avoiding the bandwidth bottleneck of moving data between subarrays and limited LUTs.
Increasing the number of LUTs can provide more parallelism in AIM, however, it significantly increases area overhead, thus, decreasing throughput-per-area. 

\emph{Sealer} has 107$\times$ and 323$\times$ lower latency than AIM-NVM for encrypting 24 and 192 data blocks, respectively.
The latency reduction for encrypting 24 data blocks comes from the architectural contribution (about 6$\times$ due to the data layout and stage fusion) and technology contribution (about 18$\times$ due to the choice of SRAM compared to NVM) in \emph{Sealer}. 
Overall, \emph{Sealer} presents a low latency encryption solution, which is suitable for real-time encryption.  

\vspace{-0.4cm}
\subsection{Throughput Normalized to Area} \label{throughput}
\begin{figure}[htp]
\vspace{-0.3cm}
\centerline{
\includegraphics[width=2.7in]{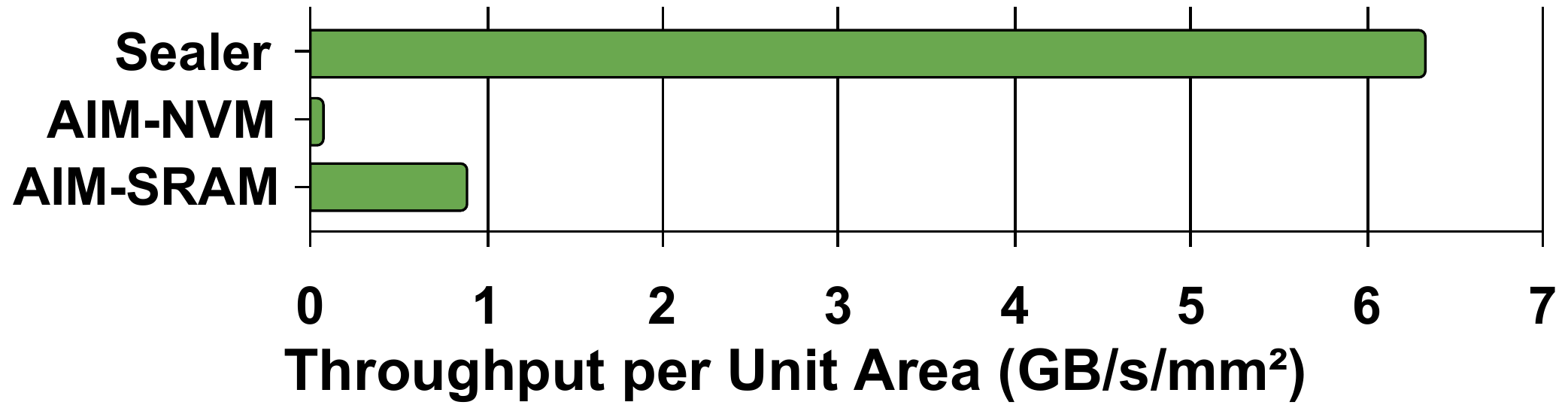}
}
\vspace{-0.3cm}
\caption{Throughput comparison among AIM-NVM, AIM-SRAM, and \emph{Sealer} for encrypting 192 data blocks.}
\label{throughput-slice}
\vspace{-0.4cm}
\end{figure}

\vspace{-0.1cm}
Figure \ref{throughput-slice} compares the throughput-per-area for \emph{Sealer}, AIM-NVM, and AIM-SRAM.
\emph{Sealer} has 7.2$\times$ higher throughput per unit area than AIM-SRAM. 
These benefits come from the fact that (1) \emph{Sealer} only incurs negligible extra overhead to the SRAM arrays by sharing the memory resources and peripheral among different stages of computation and communication, while AIM-SRAM requires additional hardware components, such as LUTs and bundles of MUX/DEMUXes, and (2) \emph{Sealer} utilizes an efficient algorithm/ architecture methodology to fuse different computational stages, which maximizes on-the-fly computation and minimizes data movement.
\emph{Sealer} can also achieve more than 91$\times$ throughput per unit area compared to the AIM-NVM. 
Although the area consumption of AIM-NVM is lower than \emph{Sealer}, the frequency of \emph{Sealer} is 133$\times$ higher than AIM-NVM; thus, resulting in higher throughput. 


\vspace{-0.3cm}
\subsection{Energy/Power Analysis}
\vspace{-0.1cm}

\begin{figure}[htp]%
\vspace{-0.5cm}
    \centering
    \subfloat{{\includegraphics[width=4cm]{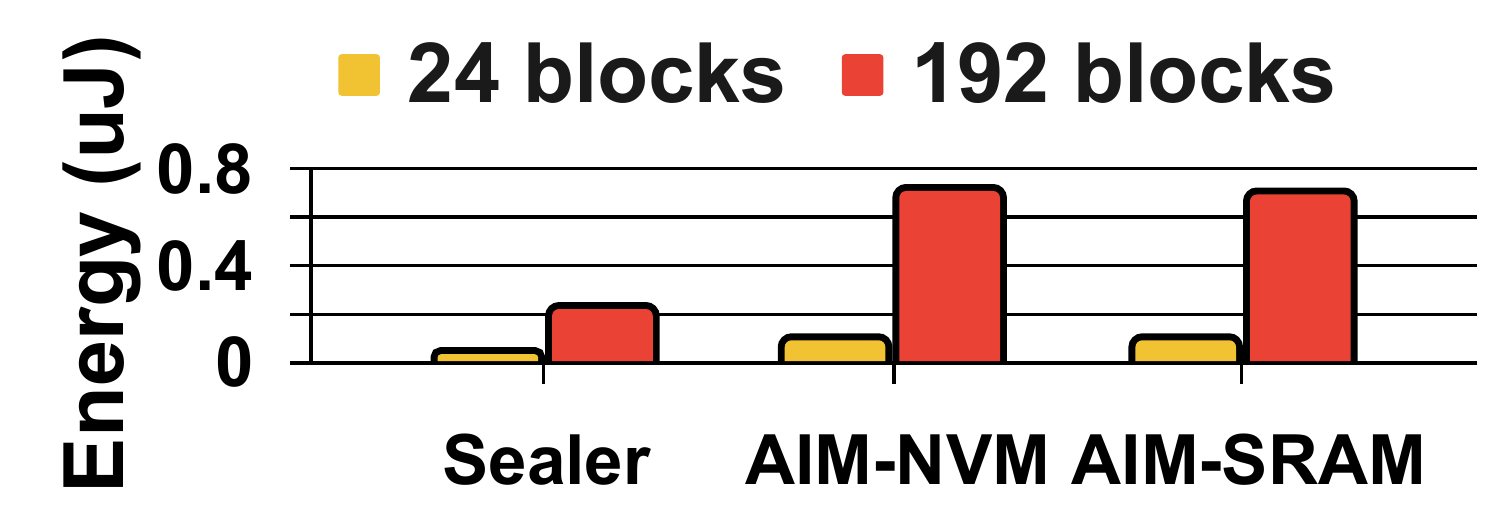} }}%
    \subfloat{{\includegraphics[width=4cm]{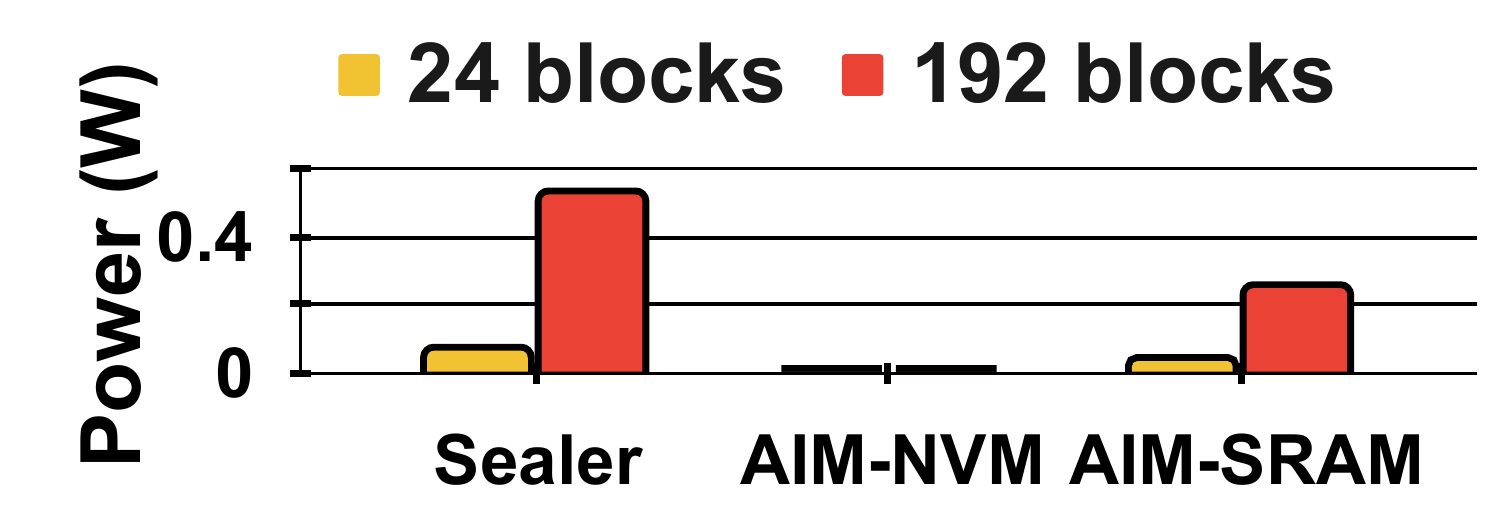} }}%
    \vspace{-0.3cm}
    \caption{Energy \& power comparison among \emph{Sealer}, AIM-NVM, and AIM-SRAM for encrypting 24 and 192 data blocks.}%
    \label{energy}%
\vspace{-0.2cm}
\end{figure}

Figure \ref{energy} compares the energy and power of Sealer, AIM-NVM, and AIM-SRAM when encrypting 24 and 192 data blocks.
To complete the encryption, \textit{Sealer} consumes 3$\times$ less energy compared to AIM-NVM and AIM-SRAM, thanks to our fusion scheme and the reduction of LUT queries.
The energy consumption of AIM-SRAM is slightly lower than AIM-NVM. This is because the \emph{MixColumns} stage, which has the highest number of operations during encryption in the AIM architecture, contains a large number of write operations (the write dynamic energy of AIM-NVM is higher than the read dynamic energy).  
Sealer has 34$\times$ and 2$\times$ higher power consumption compared to AIM-NVM and AIM-SRAM, respectively. This is because \emph{Sealer} has a higher operational frequency and also maximizes computational parallelism so that almost all units are activated and computing, which greatly increases compute/memory unit utilization. However, the other two designs have all other computing and memory units idle when querying the lookup table.

\vspace{-0.3cm}
\section{Conclusion}
\vspace{-0.1cm}

In this paper, we propose \emph{Sealer}, a real-time low-overhead in-SRAM AES engine to encrypt the plaintext on a CPU chip with narrowed trusted computing base. By efficiently mapping the algorithm to the \emph{Sealer} architecture at a finer granularity, data movement is significantly reduced.
Our evaluation results show significant performance (up to 323$\times$) and throughput-per-area (up to 91$\times$) improvement over the state-of-the-art in-memory and specialized encryption engines. Future work will extend \emph{Sealer} to provide data integrity as well as access pattern protection.
\vspace{-0.1cm}



\balance

\end{document}